\pdfoutput=1

\documentclass[12pt, hidelinks]{article}

\usepackage{import}
\usepackage[twoside=true, margin=1in]{geometry}
\usepackage{preample}

\begin{document}

\title{Credible, Strategyproof, Optimal, and Bounded Expected-Round Single-Item Auctions for all Distributions}

\author{Meryem Essaidi%
\thanks{%
    {Princeton University (\protect\url{messaidi@cs.princeton.edu}).}}
\and Matheus V.X. Ferreira%
\thanks{%
    {Harvard University (\protect\url{matheus@seas.harvard.edu}).}}
\and S. Matthew Weinberg%
\thanks{%
	{Princeton University (\protect\url{smweinberg@princeton.edu}).}}
}

\maketitle

\begin{abstract}
We consider a revenue-maximizing seller with a single item for sale to multiple buyers with i.i.d. valuations. Akbarpour and Li (2020) show that the only optimal, credible, strategyproof auction is the ascending price auction with reserves which has unbounded communication complexity. Recent work of Ferreira and Weinberg (2020) circumvents their impossibility result assuming the existence of cryptographically secure commitment schemes, and designs a two-round credible, strategyproof, optimal auction. However, their auction is only credible when buyers' valuations are MHR or $\alpha$-strongly regular: they show their auction might not be credible even when there is a single buyer drawn from a non-MHR distribution. In this work, under the same cryptographic assumptions, we identify a new single-item auction that is credible, strategyproof, revenue optimal, and terminates in constant rounds in expectation for all distributions with finite monopoly price.
\end{abstract}

\section{Introduction}

We consider a revenue-maximizing seller with a single item for sale to multiple buyers. Each buyer $i$ has value $v_i$ for the item, drawn from some distribution $D_i$ known to the seller. Our goal is to find an optimal auction among all credible auctions that are strategyproof. Informally, an auction is credible if the auctioneer has the incentive to execute the auction in earnest, even when permitted to cheat in ways that are undetectable to bidders. An auction is strategyproof/truthful if bidders have the incentive to bid their valuation $v_i$.

Akbarpour and Li \cite{AL20} introduced a trilemma for single-item auctions: (1) the second-price auction with optimal reserve is the unique truthful, one-round, revenue-maximizing auction, but is not credible; (2) the Ascending Price Auction (APA) with optimal reserve is the unique truthful, revenue-maximizing, credible auction, but requires an unbounded number of rounds of communication between the auctioneer and bidders; (3) the first-price auction with optimal reserve is the unique revenue-maximizing, credible, one-round auction, but is not truthful.

Ferreira and Weinberg \cite{ferreira2020credible} circumvent their trilemma and show that assuming the auctioneer is computationally bounded and the existence of cryptographically secure commitment schemes suffices for an optimal, strategyproof, and credible auction that terminates in two rounds---the Deferred Revelation Action (DRA). DRA derives from a simple modification of the second-price auction: initially, the auctioneer requests all bidders to cryptographically commit to their bid and submit a deposit. Later, the auctioneer requests all bidders to reveal their bids. If a bidder refuses to reveal their bid (thus aborting from the auction), their deposit is forfeit and paid to the winner of the auction. When buyers’ valuations satisfy the Monotone Hazard Rate (MHR) condition, they showed that a deposit equal to the optimal reserve price suffices to remove any incentive for the auctioneer to cheat. They also extend their analysis to $\alpha$-strongly regular valuations if one is willing to consider $\varepsilon$-credibility---the auctioneer is allowed to improve their revenue by at most an $\varepsilon$ fraction. However, the scheme doesn't extend to even a single buyer with valuation drawn from the equal revenue distribution: a cheating auctioneer can obtain infinite revenue even when the optimal auction has constant revenue! Our work proposes the Ascending Deferred Revelation Auction (ADRA) which, using dynamic deposits, extends their results to all distributions and has constant communication complexity---defined as the expected maximum number of times any bidder communicates with the auctioneer. 

The communication complexity for ADRA is a random variable that depends on a positive parameter $\varepsilon$. In the ascending price auction, while the auction has not yet ended, the auctioneer invites bidders to raise their bid in additive steps of $\varepsilon > 0$. Bidders have the option to raise their bids or drop out. The auction ends when there is a single bidder left who, in turn, pays their most recent bid. Our mechanism improves over this by using {\em multiplicative} value increments of $1+\varepsilon$ (starting from a positive reserve price $r$), which leads to an exponentially smaller communication complexity compared to APA. While APA is only approximately revenue optimal with respect to $\varepsilon$, ADRA's revenue, credibility, and strategyproofness are independent of $\varepsilon$. Thus, we can let $\varepsilon$ depend on the value distribution so that the communication complexity is constant.

Informally, our main results are (under the existence of cryptographic commitments): under the assumption bidders are drawn i.i.d. from distribution $D$ with finite monopoly price---$\arg \max_p p\pr[v \leftarrow D]{v \geq p} < \infty$---there is a truthful, revenue-maximizing, credible auction with constant communication complexity.

\paragraph{Technical Overview.} Our auctions require basic cryptographic commitments. Initially, any bidder $i$ sends a cryptographic commitment $c_i$ to a bid $b_i$, without revealing any information about their bid. The auction proceeds in rounds and in each round, we have as invariant that all bidders that have not yet quit promised that $b_i \geq q$ where $q$ is the current bid lower bound. To ensure no bidder lies (claim $b_i \geq q$ when in fact $b_i < q$), they are required to deposit a quantity $f(q) > q$ that is forfeit if they lie.

Then, in each round, the auctioneer asks each remaining bidder if $b_i \geq f(q) > q$. A bidder quits if $b_i < f(q)$, otherwise they remain in the auction. If two or more bidders raise their bid, the auction proceeds to the next round and asks bidders to increase their deposit to $f(f(q))$. Else, the auction ends and the auctioneer asks all bidders (including the ones that quit in previous rounds) to reveal their bid. The auctioneer forwards these revealed bids to all other bidders before awarding the item to the winner. If, however, a bidder does not reveal their bid or is found to have cheated in a prior round---claimed $b_i \geq f(q)$ when, in fact, $b_i < f(q)$---they lose their deposit which is paid to the winner of the auction.

Bidders' deposits have a positive externality of removing any incentive for the auctioneer to cheat in ways that bidders cannot detect. And so, the auctioneer faces a tradeoff: they can commit fake bids and selectively reveal them, but they pay a fine for every bid they choose to conceal. While a significant technical portion of ~\cite{ferreira2020credible} deals with finding a sufficiently large deposit to incentivize credibility, our work uses adaptive deposits that increase the longer a bidder remains in the auction. Furthermore, our mechanism requires constant rounds in expectation to terminate and is credible for a larger class of bidder valuations.

\paragraph{Roadmap.} Section 2 introduces notations and definitions. Section 3 defines the ascending deferred revelation auction and proves it is revenue optimal, strategyproof, and credible. Section 4 compares the communication complexity of the ascending price auction with that of our ascending deferred revelation auction.


\subsection{Related work}\label{subsec:related-work}
We contribute to a growing literature on credible auctions with the closest work to ours described in the previous section~\cite{AL20, ferreira2020credible}. Daskalakis et al.~\cite{daskalakis2020simple} proposes multi-item additive credible and approximately optimal auctions. Pycia and Raghavan \cite{raghavan2021} show that the first-price auction with no reserve price is the unique mechanism that is credible, individually rational, and that always awards the object to the highest bidder.

We also contribute to a growing literature on mechanisms with imperfect commitment in multi-period auctions. Works such as \cite{zhong2019} and \cite{skreta2013} model settings where the auctioneer is sequentially rational and cannot commit to her future behavior. In \cite{zhong2019}, the auctioneer commits to an auction with a reserve price in each period but cannot commit to her future reserve prices. In \cite{skreta2013}, the auctioneer updates her information about the buyers' values after each unsuccessful attempt to sell the item. Here, we consider an auctioneer that is unable to commit even in a single shot auction.

The use of fines to disincentivize participants from aborting a protocol is not unprecedented \cite{bentov2014,brad2008}, and there are known impossibility results when participants can abort \cite{cleve86} absent monetary incentives.

There is significant interest in using mechanism design techniques to design secure blockchains~\cite{babaioff2012bitcoin,  chen2019axiomatic, ferreira2021proof, ferreira2021dynamic, roughgarden2021transaction}. In this front, credible auction design is closely related to designing transaction fee mechanisms for blockchain where miners auction block space and (similar to here) are unable to commit to an auction format~\cite{ferreira2021dynamic, roughgarden2021transaction}.

In its inception, Bitcoin implemented a (credible) first-price auction as its transaction fee mechanism~\cite{nakamoto2008bitcoin}; however, in attempting to make transaction fees more predictable, the Ethereum Improvement Proposal (EIP) 1559~\cite{buterin2019eip1559} replaces the first-price auction with an adaptive posted-price mechanism.\footnote{EIP-1559 went live in the Ethereum blockchain in the London hard fork on August 5th, 2021~\cite{reijsbergen2021transaction}.} Roughgarden~\cite{roughgarden2021transaction} provides a formal analysis of EIP-1559. Ferreira et al. \cite{ferreira2021dynamic} propose an improvement on the pricing update rule that results in higher stability and welfare guarantees. Collusion between miners and bidders is an important concern in blockchain transaction fee mechanisms considered by \cite{roughgarden2021transaction, chung2021foundations}. In that regard, Chung and Shi~\cite{chung2021foundations} show that any strategyproof and collusion-resistant transaction fee mechanism must obtain zero revenue---the mechanism must burn all payments from bidders.

\section{Notation and Preliminaries}\label{sec:prelim}

The auctioneer has a single indivisible item to sell to $n$ bidders. Bidder $i \in [n] = \{1, 2, \ldots, n\}$ has private value $v_i$ drawn from distribution $D_i$ independently of the values of other bidders. We assume bidders have quasilinear utility: the utility of a bidder is their value (if they receive the item) minus any payments. The value profile of all bidders is the vector $\bm v = (v_1, v_2, \ldots, v_n)$ and we let $\bm D = D_1 \times \ldots \times D_n$ denote the product distribution of bidder valuations. To be concrete: each bidder $i$ knows only their value $v_i$ and type distribution $D_i$, but not the value of other bidders $\bm v_{-i} = (v_1, \ldots, v_{i-1}, v_{i+1}, \ldots, v_n)$ nor their type distributions $\bm D_{-i} = \times_{j \neq i} D_j$. When the distributions are i.i.d., we use $\bm D = D \times \ldots \times D$ to denote the product distribution of bidder valuations. The auctioneer knows the type distributions $\bm D$, but not the realization of bids $\bm v$. To sell the item, the auctioneer implements a mechanism which is a communication game between the auctioneer and the bidders. It will be convenient to refer to the communication game in its extensive form as follows.
\paragraph{Extensive-Form Game.} An {\em extensive-form game} is represented by a directed rooted tree where each node is owned by a player. We refer to the auctioneer as player $0$ and bidder $i \in [n]$ as player $i$. Each node has a set of available auctions for its owner. The owner of that node chooses one action. Upon taking this action, the game moves to the next node associated with the action taken. The game ends when the game reaches a terminal node. An {\em information set} is a collection of nodes owned by one player that are indistinguishable to that player (given all the communication up to that point).\footnote{For example, if the auction is a sealed-bid auction with two bidders, then the extensive form would have two levels. In the first level, Bidder $1$ submits a bid $b$ and we move to state $b$ where bidder $2$ is invited to submit a bid. Since bidder $2$ does not know $b$, all states $b$ are in the same information set.} 
\paragraph{The Communication Game.} We consider games that proceed as follows. In each round, the auctioneer sends a private request to each bidder which in turn sends a private reply. Whenever the auctioneer makes a request, the bidder always has the option to ignore the request and abort and leave the auction. If a bidder aborts before termination, the bidder might be required to pay a fine. To enforce fines, the auctioneer must request a bidder to make a prior deposit, and fines cannot exceed that deposit. Importantly, each bidder communicates only with the auctioneer, and upon termination learns only whether they win the item and how much they pay. The auctioneer cannot force bidders to participate honestly (or participate at all), but might use fines as described to penalize a bidder for taking certain actions.
\paragraph{Strategies.} An {\em interim strategy} $s_i$ for bidder $i$ at game $G$ is a mapping from an information set $I_i$ owned by bidder $i$ to an action $s_i(I_i)$ available at $I_i$. A {\em strategy} $S_i$ for bidder $i$ is a mapping from value $v_i$ to an interim strategy $s_{i}^{v_i}$. Given a mechanism $G$ and strategy profile $\bm S = (S_1, S_2, \ldots, S_n)$ for $G$, we refer to $(G, \bm S)$ as a protocol. $(G, \bm S)$ induces an outcome characterized by an allocation and payment rules. An {\em allocation rule} $\bm x^G$ is a vector valued function from interim strategy profile $\bm s$ to an indicator $x_i^G(\bm s) \in \{0, 1\}$ for each bidder $i \in [n]$. The indicator is 1 if and only if bidder $i$ receives the item. A {\em payment rule} $\bm p^G$ is a vector valued function from interim strategy profile $\bm s$ to the payment $p_i^G(\bm s)$ for each bidder $i \in [n]$. Then, the utility of bidder $i$ is
$$u_i^G(\bm s) := v_i \cdot x_i^G(\bm s) - p_i^G(\bm s).$$
When $(G, \bm S)$ is clear from context, we write $\bm x$, $\bm p$, $\bm u$ to denote the allocation, payment rule, and utility respectively. The payoff the auctioneer derives from $(G, \bm S)$ is the revenue:
$$\rev^{(G, \bm S)}(\bm D) = \e[\bm{v} \leftarrow \bm{D}]{\sum_{i = 1}^n p_i(\bm s)}.$$
\paragraph{Equilibrium.} An interim strategy $S_i(v_i)$ is a {\em best response} to interim strategy $\bm S_{-i}(\bm v_{-i})$ of everyone else if bidder $i$ (weakly) maximizes their utility by following strategy $S_i(v_i)$ (over any other strategy $S_i'(v_i)$). Our first desiderata is to design mechanisms that have an {\em ex-post Nash Equilibrium}. Formally, $(G, \bm S)$ forms an ex-post Nash equilibrium if for all $i \in [n]$ and for all $\bm v$, $S_i(v_i)$ is a best response to $\bm S_{-i}(\bm v_{-i})$.
\begin{definition}[Strategyproof/IC]
Mechanism $G$ is {\em ex-post Incentive Compatible} (IC) if there is a strategy profile $\bm S$ such that $(G, \bm S)$ forms an ex-post Nash equilibrium.
\end{definition}
The existence of a Bayesian Nash equilibrium is weaker than that of an {\em ex-post} Nash equilibrium. The seminal work of Myerson~\cite{myerson1981optimal} characterizes the expected revenue of any mechanism at a Bayesian equilibrium.
\begin{definition}
$(G, \bm S)$ forms a Bayesian Nash equilibrium if for all $i \in [n]$, for all $v_i$,
$$S_i(v_i) \in \arg\max_{S_i'(v_i)} \e[\bm{v_{-i}} \leftarrow \bm{D_{-i}}]{u_i(v_i, S_i'(v_i), \bm S_{-i}(\bm v_{-i}))}.$$
Moreover, $G$ is {\em Bayesian Incentive compatible} (BIC) if there is a strategy profile $\bm S$ such that $(G, \bm S)$ forms a Bayesian Nash equilibrium.
\end{definition}
\begin{definition}[Virtual Value]
The {\em Cumulative Density Function} (CDF) of distribution $D$ is $F(x) = \pr[v \leftarrow D]{v \leq x}$ and the {\em Probability Density Function} (PDF) of $D$ is $f(x):=dF(x)/dx$. The {\em virtual value} function of $D$ is the real-valued function $\vvalue^D(x) = x-\frac{1-F(x)}{f(x)}$. The {\em ironed virtual value} $\ivvalue$ of $D$ is the upper concave envelope of $\vvalue$ (see~\cite{myerson1981optimal}). We will often abuse notation and use $\vvalue_i(x) := \vvalue^{D_i}(x)$ and $\ivvalue_i(x):=\ivvalue^{D_i}(x)$.
\end{definition}
\begin{theorem}[Myerson'81 \cite{myerson1981optimal}]\label{thm:myerson}
Assume $G$ is BIC and bidders follow Bayesian equilibrium $\bm S$. Then
$$\rev^{(G, \bm S)}(\bm D) = \e[\bm v \leftarrow \bm D]{\sum_{i = 1}^n \ivvalue_i(v_i) \cdot x_i(\bm s)}.$$
\end{theorem}
We refer to $\rev(\bm D)$ as the optimal revenue over all BIC mechanisms at a Bayesian equilibrium. If the virtual value function $\vvalue_i(\cdot)$ is monotone non-decreasing, we say the distribution $D_i$ is regular:
\begin{definition}[$\alpha$-strongly regular/Regular/MHR]
A distribution $D$ is {\em $\alpha$-strongly regular} if for all $v' \geq v$, $\vvalue^D(v') - \vvalue^D(v) \geq \alpha (v' - v)$. Moreover, $D$ is {\em regular} if $D$ is $0$-strongly regular and $D$ has {\em Monotone Hazard Rate} (MHR) if $D$ is $1$-strongly regular.
\end{definition}
Thus if $D_i$ is regular, for all $i$, the optimal auction allocates the item to the bidder with the highest virtual value, as long as their virtual value is non-negative. This defines an optimal reserve price $r(D) := (\ivvalue^D)^{-1}(0)$, commonly referred as the {\em Myerson reserve} for distribution $D$. Note the ironed virtual value $\ivvalue$ function is non-decreasing. Moreover, we will only consider distributions where the Myerson reserve is finite, which is equivalent to the monopoly price $\arg\max_p p\pr[v \leftarrow D]{v \geq p}$ being finite.

Our second desiderata concerns credibility---the incentive for the auctioneer to implement the promised auction. While the auctioneer promises to implement a mechanism $G$, we assume the auctioneer might deviate as long as this deviation cannot be detected by bidders. To be concrete: the auctioneer and bidders participate in a protocol $(G, \bm S)$ and a {\em safe deviation} of the auctioneer's strategy is another mechanism $G'$ such that no bidder can distinguish the protocol $(G', \bm S)$ from $(G, \bm S)$. This is because their personal communication with the auctioneer is {\em always} consistent with the auctioneer implementing the intended mechanism $G$ to {\em some} set of $n_i$ bidders, who are using feasible strategies with {\em some} value profile $v_i^1, v_i^2, \ldots, v_i^{n_i}$ (we add the subscript $i$ to denote that the set of bidders need not be the same for all $i$, and the hypothetical strategies need not be the same either). Importantly, because each bidder can only observe their own communication with the auctioneer, distinct bidders can have inconsistent views of what the communication game is, depending on their interaction with the auctioneer. Then, although the auctioneer is playing protocol $(G', \bm S)$, bidder $i$ believes the auctioneer is playing the protocol $(G, S_i, \bm{(S_i)}_{-i})$ where $\bm{(S_i)}_{-i}$ is any strategy profile for $n_i$ bidders.

Before we formally define credibility, we will require the same computational assumptions from \cite{ferreira2020credible}.
\begin{definition}\label{def:crypto} A {\em commitment scheme} $\commit$ with parameter $\lambda$ is an algorithm that takes a message $m \in \{0, 1\}^{\poly(\lambda)}$ and a random string $r \in \{0, 1\}^{\poly(\lambda)}$ and outputs a commitment $\commit(m, r)$ and satisfy the following conditions:
\begin{itemize}
    \item \textbf{Efficiency.} Evaluating $\commit(m, r)$ takes time $\poly(\lambda)$.
    \item \textbf{Perfectly Hiding.} The distributions of $\commit(m, r)$ and $\commit(m', r')$, when $r$ and $r'$ are uniformly random, are identical for any messages $m$ and $m'$.
    \item \textbf{Computationally Binding.} For any algorithm $A$ that takes as parameter $\lambda$, the probability $A$ outputs pairs $(m, r) \neq (m', r')$ such that $\commit(m, r) = \commit(m', r')$ is at most $\frac{1}{2^{\Omega(\lambda)}}$.
    \item \textbf{Non-malleable.} Formal definitions of malleability are involved (see \cite{dolev2003nonmalleable}). At a high level, a commitment scheme is {\em non-malleable} if given a commitment $c = \commit(b, r)$ and a non-identity function $f$ and $g : \{0, 1\}^* \to \{0, 1\}^*$, it is not possible to compute a commitment $\commit(f(b), g(r))$.
\end{itemize}
\end{definition}
\begin{definition}[Reasonable deviation]
A commitment $c$ is {\em explicitly tied} to $(b, r)$ if either the auctioneer or a bidder computed $c = \commit(b, r)$. A deviation for the auctioneer is {\em reasonable}, if whenever the auctioneer opens a commitment $c = \commit(b, r)$---by revealing $(b, r)$---then $c$ was explicitly tied to $(b, r)$.
\end{definition}
Note the assumption that a commitment scheme is computationally binding implies a commitment $c$ is explicitly tied to at most one pair $(b, r)$. Non-malleability is important to motivate the restriction of the auctioneer to reasonable deviations only: if our commitment scheme were malleable, it would be possible for the auctioneer to reveal a commitment $c = \commit(f(b), g(r))$ without ever explicitly computing $\commit(f(b), g(r))$---they would observe $\commit(b, r)$, indirectly compute a commitment $\commit(f(b), g(r))$ and reveal $f(b)$ once they learn $(b, r)$.
\begin{definition}[Computationally Credible]
For a mechanism $G$, assume bidders follow a strategy profile $\bm S$ where $(G, \bm S)$ is a Bayesian Nash equilibrium. $(G, \bm S)$ is {\em computationally credible} if implementing the communication game $G$ maximizes the auctioneer's expected revenue over all safe and reasonable deviation where the expectation is taken over bidder valuations.
\end{definition}
Our last desiderata concerns minimizing the number of times a bidder is required to communicate with the auctioneer.
\begin{definition}[Communication/Round Complexity]
The {\em communication/round complexity} of protocol $(G, \bm S)$ is the expected maximum number of times any bidder communicates with the auctioneer.
\end{definition}
As an example, the communication complexity for direct revelation auctions (like the first and second price auctions) is 1; while the communication complexity for the Deferred Revelation Auction (DRA) from \cite{ferreira2020credible} is 2.

\section{Ascending Deferred Revelation Auction}\label{sec:adra}

Let's first recall a formal definition of the Ascending Price Auction (APA).
\begin{definition}[Ascending Price Auction]\label{def:apa}
The {\em ascending price auction} (APA) with reserve price $r$ and positive step size $\varepsilon$ proceeds as follows:
\begin{enumerate}
\item Initially, bidder $i$ bids $b_i = r$; otherwise, they quit the auction.
\item While there are two or more bidders that did not quit:
\begin{enumerate}
\item \label{apa:3} The auctioneer visits the bidder with lowest bid. Let $i$ be such bidder. The auctioneer asks bidder $i$ if they wish to raise their bid by $\varepsilon$. If they accept, update their bid to $b_i = b_i + \varepsilon$. Else, bidder $i$ quits the auction, does not receive the item, and pays $0$.
\end{enumerate}
\item Allocate the item to the (unique) bidder $i^*$ that did not quit. Bidder $i^*$ pays their bid $b_{i^*}$.
\end{enumerate}
\end{definition}
From \cite{AL20}, APA is the only optimal, credible, strategyproof mechanism; but its expected number of rounds to terminate can be large:
\begin{lemma}\label{lem:apa-lowerbound}
Consider the ascending price auction with $n$ bidders with valuations drawn from distribution $\bm D$ where bidder $i$ quits when asked to bid above $v_i$. Then the communication complexity for APA is  $\Theta(1) + \frac{\rev^{(G, \bm S)}(\bm D) - r\pr{\max_{i}v_i \geq r}}{\varepsilon}$.
\end{lemma}
We defer the proof to Appendix~\ref{app:adra-proofs}. Intuitively, APA requests bidders to raise bids in {\em additive} value increments until there is at most one bidder left. Note that the ascending price auction with Myerson reserve is approximately revenue optimal: its revenue approximates the optimal revenue as $\varepsilon \to 0$, which, in turn, increases the communication complexity. The next example provides an application of Lemma~\ref{lem:apa-lowerbound}.
\begin{example}\label{example:equal-revenue}
Consider APA with reserve price $r = 1$ and $n$ bidders drawn from the equal revenue distribution $D$ with $\pr[v \leftarrow D]{v \geq p} = \frac{1}{p}$. From Theorem~\ref{thm:myerson}, APA with reserve price $r = 1$ is revenue optimal with $\rev(D^n) = n$ (see Appendix~\ref{app:adra-proofs} for a proof). Then Lemma~\ref{lem:apa-lowerbound} implies that the communication complexity on this instance is $\Theta(\frac{n}{\varepsilon})$.
\end{example}
The Deferred Revelation Auction (DRA) from \cite{ferreira2020credible} improves upon APA by requiring only two rounds of communication between each bidder with the auctioneer. However, the DRA is only credible when bidder valuations are drawn from an MHR distribution. This MHR condition is not satisfied by the equal revenue distribution from Example~\ref{example:equal-revenue}: even if there is a single bidder, for any positive $M > 0$ and any penalty $P > 0$, \cite{ferreira2020credible} showed a safe deviation for that auctioneer that obtains revenue at least $M$ despite the fact that the optimal auction has revenue 1!

To circumvent the limitations of DRA and APA, we propose an intermediate mechanism, the Ascending Deferred Revelation Auction (ADRA), that will combine properties from both auctions. It proceeds in a similar fashion to APA. In each round, if a bidder is still in the auction and is bidding $b$, the auctioneer asks if that bidder wishes to increase their bid to $f(b) > b$ where $f$ is any non-decreasing function. For APA, $f(b) = b + \varepsilon$. For ADRA, we let $f$ be an arbitrary increasing function and in particular, we study the communication complexity when $f(b) = (1+\varepsilon) \cdot b$---bids increase in multiplicative rather than additive steps.

For ADRA, if a bidder refuses to raise their bid, they {\em tentatively quit} from the auction, but they will still have the chance to receive the item. Note the distinction from APA: in APA, whenever a bidder quits, they have no chance of receiving the item and pay nothing. This ensures that APA is credible: if the auctioneer lies and attempts to make the highest bidder increase their bid when they are the only bidder left, the auctioneer risks not allocating the item to anyone.

For ADRA, all bidders are required to commit to their bid before the auction starts and submit a deposit (just like in DRA), but the deposit increases the longer a bidder stays in the auction (unlike DRA where the deposit is fixed). This allows ADRA to be credible in instances where DRA is not. Also commitments schemes allows ADRA to be credible even if a bidder have a second chance to receive the item (after they quit).

During the execution of the auction, bidders must behave according to their committed bid. That is, if bidder $i$ commits to bid $b_i$, then they must raise their bid as long as their bid does not exceed $b_i$, and they must quit if their bid is about to exceed $b_i$. Whenever a bidder raises their bid, they also raise their deposit. If a bidder deviates (refuses to participate, or raises their bid when $f(b) > b_i$, or quits when $f(b) < b_i$), they will forfeit their deposit to the winner of the auction. In general, no honest bidder is required to deposit more than $f(b_i) = O(b_i)$. Next, we proceed with a formal definition of ADRA.
\begin{definition}[Levels]
For each bid $b$, we assign an integer value $g(b)$ where $g$ is non-decreasing---for two bids $b \neq b'$,  $b > b'$ if and only if$g(b) \geq g(b')$. We refer to $g$ as a {\em level function} and $g(b)$ as the {\em level} of $b$. We define the inverse function $g^{-1}(k) := \sup\{x : g(x) = k\}$. Note $g^{-1}$ is non-decreasing.
\end{definition}
\begin{definition}[Ascending Deferred Revelation Auction]\label{def:adra}
The {\em Ascending Deferred Revelation Auction} (ADRA) with reserve price $r$ and level function $g$, proceeds as follows:
\begin{enumerate}
\item We refer to $d_i$ as the {\em deposit of bidder $i$}. Initially $d_i = r$. During execution of the auction, we say a bidder {\em aborts} whenever they refuse to follow with the auctioneer's request. Let $\ell_i$ denote the level bidder $i$ quits, aborts, or becomes the only remaining bidder. Let $k$ be the current level of the auction. Initialize $k := g(r)$. 

\item \label{step:adra-commit} Request bidder $i$ to commit to a bid $b_i$---bidder $i$ draws a uniformly random string $r_i$ and sends $c_i = \commit(b_i, r_i)$ to the auctioneer.

\item The auctioneer forwards $c_i$ to bidder $j \neq i$.

\item \label{step:adra-while} While there are two or more bidders that have neither quit nor aborted, for each bidder $i$ that is still active:

\begin{enumerate}
    \item \label{step:adra-level} If $g(b_i) \leq k$, request bidder $i$ to quit.
    
    \item If $g(b_i) > k$, request bidder $i$ to raise their deposit to $d_i = g^{-1}(k+1)$.
    
	\item Increment $k = k + 1$ and return to Step~\ref{step:adra-while}.
\end{enumerate}

\item \label{step:adra-reveal} Request bidder $i$ to reveal $(r_i, b_i)$. The auctioneer forwards $(r_i, b_i)$ to bidder $j \neq i$.

\item The auctioneer aborts bidder $i$ if:
	\begin{enumerate}
	\item Bidder $i$ refuses to reveal $(r_i, b_i)$, or
	\item Bidder $i$ sends a pair $(r_i', b_i')$ such that $c_i \neq \commit(r_i', b_i')$, or
	\item Bidder $i$ correctly sends a pair $(r_i, b_i)$ such that $c_i = \commit(r_i, b_i)$, but, at Step~\ref{step:adra-level}, bidder $i$ did not quit at the first level $\ell_i$ where $g(b_i) \leq \ell_i$.
	\end{enumerate}

\item Let $A$ denote the set of bidders that did not abort. Let $i^* = \arg\max_{i \in A : b_i \geq r} b_i$ be the highest such bidder (if one exist). Bidder $i^*$ receives the item and pays
$$p_{i^*} = \max\left\{r, \max_{j \in A \setminus \{i^*\}} b_j\right\}.$$
\item Bidder $i \in A$ receives $d_i$ back. Bidder $i^*$ receives the deposit of bidder $j \not \in A$.
\end{enumerate}
\end{definition}
There are many safe and reasonable deviations that the auctioneer can implement:
\begin{itemize}
    \item At Step~\ref{step:adra-commit}, commit to a bid $b \neq b_j$, for any $j \in [n]$ to bidder $i$.
    
    \item Commit to a bid $b$ to bidder $i$ once they learn about bid $b_j$---$b$ can depend on $b_j$, but not $b_i$ since the commitment scheme is non-malleable and perfectly hiding.
    
    \item Commit to a bid $b$ to bidder $i$, but not to bidder $j$.
    
    \item Commit to a bid $b$ to bidder $i$, but not reveal $b$ at Step~\ref{step:adra-reveal}.
\end{itemize}
ADRA instructs bidders to follow the following strategy:
\begin{definition}
Assume bidder $i$ participates in the ascending deferred revelation auction with reserve price $r$ and level function $g$. Define $S_i^{ADRA}(b_i)$ as the {\em suggested strategy} for bidder $i$ where, given bid $b_i$ (not necessarily equals to $v_i$), bidder $i$:
\begin{itemize}
    \item Commits to bid $b_i$ by forwarding $c_i = \commit(b_i, r_i)$ where $r_i$ is a uniformly random string;
    \item Quits at the first level $k$ where $k \leq g(b_i)$ (recall $k \geq g(r)$);
    \item Reveals $(b_i, r_i)$ once there is a single bidder left.
\end{itemize}
We say bidder $i$ is {\em truthful} when they implement $S_i^{ADRA}(v_i)$.
\end{definition}
Next, we observe that implementing the suggested strategy is a dominant strategy for bidders (even when the auctioneer cheats); otherwise, that bidder loses their deposit.
\begin{observation}\label{obs:bidder-strategy}
Assume the auctioneer implements any safe deviation from ADRA. Then it is a best response for bidder $i$ to implement the suggested $S_i^{ADRA}(b_i)$ for some bid $b_i \leq v_i$.
\end{observation}
\begin{proof}
If bidder $i$ deviates from $S_i^{ADRA}$, they abort from the auction, do not receive the item, and have their deposit forfeit. By implementing $S_i^{ADRA}(b_i)$, bidder $i$ either wins and pay at most $v_i$ or loses and pay nothing. Thus, following $S_i^{ADRA}(b_i)$, for some $b_i \leq v_i$, weakly dominates any other strategy.
\end{proof}
Next, we prove that the ADRA with optimal reserve price is credible. The proof follows a similar format to that of the proof of credibility for the ascending price auction with optimal reserve~\cite{AL20}. \cite{AL20} observes the only safe deviation for the auctioneer is to force the highest bidder to increase their bid even when they are the only bidder left in the auction. However, even if the auctioneer cheats, it is a dominant strategy for the highest bidder to stay in the auction as long as their bid does not exceed their value $v_i$, since they receive nothing once they quit. This argument implies any safe deviation from APA is itself a BIC mechanism---bidders would not have the incentive to change their strategy even if they knew the auctioneer was cheating. As a result, no safe deviation (which is a BIC mechanism) can provide higher revenue than the ascending price auction with optimal reserve (which is a revenue optimal BIC mechanism). This proves the ascending price auction with optimal reserve is credible.\footnote{Credibility is a subtle property. An innocent modification of the ascending price auction (which is not credible) is to allow the auctioneer to simultaneously request all bidders to raise their bid by $\varepsilon$. In case all remaining bidders quit simultaneously, the auctioneer allocates the item to a random bidder and charges their previous bid. Despite the similarities and the fact this auction implements the same allocation and payment rules as the ascending price auction, this implementation is not credible since there is a safe deviation which forces the highest bidder to quit when their bid would exceed $\max v_i$ and charge at least $\max v_i - \varepsilon$: the auctioneer claims there is another bidder in the auction until the highest bidder quits, at which point, the auctioneer claims the other bidder left at the same time giving the item to bidder $i$. This happens because (differently from the ascending price auction) a bidder can still receive the item once they quit.}

To apply a similar argument to ADRA, we note that the strategy profile $\bm S^{ADRA}(\bm v) = (S_1^{ADRA}(v_1), S_2^{ADRA}(v_2), \ldots, S_n^{ADRA}(v_n))$ for ADRA forms an ex post Nash equilibrium: it results in the same allocation and payments as the second-price auction with reserve $r$. We will show that $\bm S^{ADRA}(\bm v)$ still forms an ex post Nash equilibrium even if the auctioneer implements an optimal safe deviation. Note the distinction with Observation~\ref{obs:bidder-strategy} which doesn't specify the value for $b_i$, while here we will explicitly claim that setting $b_i = v_i$ and implementing $S_i^{ADRA}(v_i)$, for all $i$, forms an ex post Nash equilibrium.

As a result, any safe deviation of ADRA is a BIC mechanism---bidder $i$ is better of being truthful assuming bidder $j \neq i$ is truthful. Thus if ADRA is already revenue optimal (among all BIC mechanisms), the auctioneer cannot improve revenue by implementing a safe deviation (which is itself a BIC mechanism).
\begin{proposition}\label{prop:adra-nash}
Assume bidder $i$ implements $S_i^{ADRA}(v_i)$. If $G'$ is a safe deviation from ADRA, then $(G', \bm S^{ADRA})$ forms an {\em ex post} Nash equilibrium.
\end{proposition}
We defer the proof to the appendix. For intuition, consider the ADRA with any level function $g$, and a single bidder with value $v$ drawn from the equal revenue distribution. The optimal auction sets a reserve price $r = 1$ resulting in a revenue of 1. We argue that the auctioneer cannot improve revenue by cheating---while the auctioneer could obtain infinite revenue under DRA~\cite{ferreira2020credible}. 

Informally, if the auctioneer deviates in the attempt to force the real bidder to pay more than 1, they would impersonate fake bidders and choose which bids to reveal during the execution of the auction. However, the auctioneer must abort any fake bidders that bid above $v$; otherwise, the real bidder knows they are not the winner. The challenge (for the auctioneer) is that the longer a fake bidder remains active in the auction, the more their deposit increases (thus making it more expensive to abort). We argue that the real bidder cannot improve their utility by bidding $b \neq v$ under this safe deviation. For simplicity, consider the case where the auctioneer sends only one fake bid $b'$ independent of $b$ (since our commitment scheme is non-malleable and perfectly hiding).

The case where the real bidder bids $b > v$ can only result in a higher payment than $v$ when $b' > v$ and the same outcome when $b' < v$. For the case where the real bidder bids $b < v$, if $b' \not \in (b, v)$, the outcome is the same regardless of the real bidder bidding $b$ or $v$. However, if $b' \in (b, v)$, the auctioneer could make the real bidder pay $b'$ when the real bidder bids $v$. The catch is that the auctioneer cannot distinguish the cases where the real bidder bids $v$ or $b$ until the real bidder quits at level $g(b)$---since our commitment scheme is perfectly hiding. To be concrete: if the auctioneer aborts $b'$ by level $g(b)$, they would have done so before knowing if the real bidder bids $b$ or $v$ which also results in the same outcome regardless of the real bidder bidding $b$ or $v$. However, if the auctioneer does not abort $b'$ by level $g(b)$, they would not abort $b'$ once the real bidder quits at level $g(b)$ since the payment the auctioneer receives is at most $b$ and their fines for aborting $b'$ is at least $g^{-1}(g(b)) \geq b$. Thus, the auctioneer reveals $b' > b$ and the real bidder does not receive the item. Informally, this proves the real bidder is better off by bidding $v$ even if the auctioneer cheats.
\begin{theorem}\label{thm:adra-credible}
For any reserve price $r$ and level function $g$, ADRA is strategyproof. Moreover, assume bidder valuations are drawn i.i.d. from distribution $D$. Then ADRA with reserve price $r(D)$ is revenue-optimal and computationally credible.
\end{theorem}
\begin{proof}
Strategyproofness follows directly from Proposition~\ref{prop:adra-nash}: if the auctioneer commits to implement ADRA, the strategy profile $\bm S^{ADRA}(\bm v)$ forms an {\em ex post} Nash equilibrium.

Let's check that ADRA is revenue optimal. Observe that ADRA maximizes virtual welfare, since it only allocates the item to the highest bidder if their ironed virtual value is non-negative. Indeed, $v_{i^*} \geq r(D)$ if and only if $\ivvalue(v_{i^*}) \geq \ivvalue(r(D)) = 0$. From Theorem~\ref{thm:myerson}, ADRA is revenue optimal.

To check that ADRA with Myerson reserve is computationally credible, recall that Proposition~\ref{prop:adra-nash} states that any safe deviation from ADRA is itself a BIC mechanism. Since ADRA with reserve $r(D)$ is a revenue optimal BIC mechanism, no safe deviation can provide more revenue than honestly implementing ADRA. This proves that ADRA is computationally credible as desired.
\end{proof}

\paragraph{Is ADRA DSIC?} One might wonder if truthful bidding---bidding $b_i = v_i$---is {\em Dominant Strategy Incentive Compatible} (DSIC)---a stronger equilibrium guarantee than {\em ex post} IC. To be concrete, a protocol $(G, \bm S)$ is DSIC if for all $i \in [n]$, for all $\bm v$, and for all $\bm S_{-i}'$,
$$S_i(v_i) \in \arg\max_{s_i'} u_i(v_i, s_i, \bm S_{-i}'(\bm v_{-i})).$$
That is, bidder $i$ prefers to play $S_i$ regardless of the strategy of other bidders. Unfortunately, that is not the case. The problem is that, during ADRA's execution, bidder $j \neq i$ receives information about bidder $i$'s strategy at every level---and bidder $i$ must best-respond even to unnatural strategies from bidder $j$. As an example, consider a strategy where bidder $j \neq i$ implements $S_j^{ADRA}(v_j)$ unless bidder $i$ quits at level $g(r)$ in which case bidder $j$ aborts at level $g(r)$. Thus for bidder $i$, it is optimal to bid $b_i = r$ and quit at level $g(r)$ so they win the item and pay $r$. The behavior of bidder $j \neq i$ is sub-optimal but highlights that DSIC is a very strong condition for our setting.\footnote{The ascending price auction is not only DSIC, but it also satisfy an even stronger guarantee known as obviously strategyproofness~\cite{li2017obviously}: if your next bid is not going to exceed your value, the best outcome for quitting is not better than the worst outcome for raising your bid and quitting next round (thus there is no risk for raising your bid).}

\section{Communication Complexity}\label{sec:communication}

In this section, we study the improved communication complexity of the Ascending Deferred Revelation Auction (ADRA) over the Ascending Price Auction (APA). When setting the level function as $g(b) = \lceil\log_{1+\varepsilon}(b/r)\rceil$, we obtain an exponential improvement on the communication complexity compared with APA for fixed $\varepsilon$. Moreover, by letting $\varepsilon$ depend on the distribution $\bm D$, ADRA can have constant communication complexity.
\begin{theorem}\label{thm:adra-upperbound}
Consider the ascending deferred revelation auction with reserve price $r$, level function $g(b) = \lceil\log_{1+\varepsilon}(b/r)\rceil$, and $n$ bidders drawn from distribution $\bm D$, where bidder $i$ bids $v_i$ and quits at the first level $k$ where $k \geq g(v_i)$. Then the communication complexity is $O(\frac{\log_{1+\varepsilon}(\rev^{(G, \bm S)}(\bm D))}{r})$.
\end{theorem}
We defer the proof to Appendix~\ref{app:adra-proofs}. Intuitively, our choice for $g$ provides {\em multiplicative} value increments until there is at most one bidder left. This provides communication complexity that is logarithmic with respect to the revenue of the auction. Differently from APA, our communication complexity bounds are also scale-invariant---multiplying all values and the reserve price by a constant $c$ results in the same communication complexity. The following example shows that Theorem~\ref{thm:adra-upperbound} provides a tight bound:


\begin{example}
Consider the ascending deferred revelation auction with reserve price $r = 1$, level function $g(b) = \lceil\log_{1+\varepsilon}(b/r)\rceil$, and $n \geq 2$ bidders drawn i.i.d. from the equal revenue distribution $D$---$\pr[v \leftarrow D]{v > x} = \frac{1}{x}$. From Example~\ref{example:equal-revenue}, the revenue for this auction is $\rev(\bm D) = n$ and from Theorem~\ref{thm:adra-upperbound}, we find that the communication complexity for ADRA is at most $O(\log_{1+\varepsilon}(n))$. Let $x = \inf\{y : \pr[v \leftarrow D]{v > y} \geq \frac{1}{n}\}$. Let $u(\bm v)$ be the second highest bid among $v_1, v_2, \ldots, v_n$.
Since $(1-\frac{1}{n})^n \leq \frac{1}{e}$ for any $n\geq 1$, we find that $\pr{u(\bm v) > x} = 1 - \pr{u(\bm v) \leq x} \geq 1 - \left(1-\frac{1}{n}\right)^{n-1} \geq 1 - \frac{(1-\frac{1}{n})}{e} \geq 1 - \frac{2}{e} $.
And so, with constant probability, the second-highest bidder has value at least $n$ which implies the auction takes at least $\log_{1+\varepsilon} n$ rounds to terminate. This shows that Theorem~\ref{thm:adra-upperbound} is tight. Setting $\varepsilon = n$ results in an auction with constant communication complexity.
\end{example}
Next, we show the communication complexity explicitly in terms of the cumulative density function; however, we will require the assumption values are drawn i.i.d. from a regular distribution. Intuitively, we first define a value $x$ large enough to exceed most bids with high probability. As a consequence, most bidders drop out before level $x$, after which, only a constant expected number of rounds is required until termination. If ADRA further sets its multiplicative step size $\varepsilon$ to be $x$, the auction has constant communication complexity.
\begin{theorem}\label{thm:regular}
Let $D$ be a regular distribution. Consider the ADRA with optimal reserve price $r(D)$, level function $g(b) = \lceil\log_{1+\varepsilon}(b/r(D))\rceil$, and $n$ bidders with valuations drawn i.i.d. from $D$ where bidder $i$ bids $v_i$ and quits at the first level $k$ where $k \geq g(v_i)$. Then the communication complexity is $O(\max\{\frac{1+\varepsilon}{\varepsilon},\log_{1+\varepsilon}\left(\frac{x}{r(D)}\right)\})$ where $x \coloneqq \inf\{p : \pr[v \leftarrow D]{v \geq p} \leq \frac{1}{n}\}$.
\end{theorem}
Note that for some constant $c > 0$, setting $\varepsilon = \min\{c, x\}$ ensures the communication complexity for ADRA is constant. Next, we consider an application of Theorem~\ref{thm:regular}.
\begin{example}
Consider ADRA with reserve price $r = 1$, level function $g(b) = \lceil\log_{1+\varepsilon}(b/r)\rceil$, and $n$ bidders with valuations drawn i.i.d from the exponential distribution $D$---$\pr[v \leftarrow D]{v > p} = e^{-p}$. Because $D$ is regular, Theorem~\ref{thm:regular} provides a direct route to see the communication complexity is $O(\log\log(n))$. For that observe $x = \inf\{p : \pr[v \leftarrow D]{v \geq p} \leq \frac{1}{n}\} = F^{-1}(1-\frac{1}{n}) = \log(n)$. Consequently, from Theorem~\ref{thm:regular}, the communication complexity is $O(\log_{1+\varepsilon}x) = O(\log_{1+\varepsilon}\log(n))$. Setting $\varepsilon = \log(n)$ results in an optimal, credible, truthful auction with constant communication complexity.
\end{example}

\section{Conclusion}\label{sec:conclusion}

We extend the work of Ferreira and Weinberg \cite{ferreira2020credible} on credibility for single-item auctions in two ways: we introduce a mechanism that generalizes to any distribution $D$ with finite monopoly price. Our main technical result is that the Ascending Deferred Revelation Auction (ADRA) is credible, strategyproof, revenue optimal, and terminates in constant rounds in expectation.

For a fixed $\varepsilon > 0$, using \emph{multiplicative} bid increments of $(1 + \varepsilon)$, ADRA provides an exponential improvement on the communication complexity of the Ascending Price Auction (APA) which uses {\em additive} bid increments of $\varepsilon$. In particular, while APA has communication complexity proportional to the revenue $\rev(\bm D)$, ADRA has communication complexity proportional to $\log\left(\frac{\rev(\bm D)}{r}\right)$ where $r$ is the reserve price. Setting $\varepsilon = \varepsilon(n, \bm D)$ as a direct function of the number of bidders $n$ and value distributions $\bm D$, we identify an instance of ADRA with constant communication complexity. For instances where bidder valuations are drawn i.i.d. from a regular distribution $D$, we provide communication complexity bounds that depend only on the cumulative density function of $D$. As an application, setting $\varepsilon = \log\log(n)$ and using Myerson reserve price results in a credible, truthful, (optimal) auction that terminates in a constant number of rounds when valuations are drawn independently from the exponential distribution.

To overcome the limitations of the Deferred Revelation Auction (DRA), ADRA combines cryptographic commitments with dynamic deposits. Reducing the required deposits is an interesting open question that has practical implications; although, ADRA requires no bidder to deposit more than $(1+\varepsilon)$ fraction of their valuation. Another direction is to extend the credibility of ADRA beyond the i.i.d. framework.

\bibliographystyle{plainnat}
\bibliography{mybib}

\begin{thebibliography}{21}
\providecommand{\natexlab}[1]{#1}
\providecommand{\url}[1]{\texttt{#1}}
\expandafter\ifx\csname urlstyle\endcsname\relax
  \providecommand{\doi}[1]{doi: #1}\else
  \providecommand{\doi}{doi: \begingroup \urlstyle{rm}\Url}\fi

\bibitem[Akbarpour and Li(2020)]{AL20}
Mohammad Akbarpour and Shengwu Li.
\newblock Credible auctions: A trilemma.
\newblock \emph{Econometrica}, 88\penalty0 (2):\penalty0 425--467, 2020.

\bibitem[Babaioff et~al.(2012)Babaioff, Dobzinski, Oren, and
  Zohar]{babaioff2012bitcoin}
Moshe Babaioff, Shahar Dobzinski, Sigal Oren, and Aviv Zohar.
\newblock On bitcoin and red balloons.
\newblock In \emph{Proceedings of the 13th ACM conference on electronic
  commerce}, pages 56--73, 2012.

\bibitem[Bentov and Kumaresan(2014)]{bentov2014}
Iddo Bentov and Ranjit Kumaresan.
\newblock How to use bitcoin to design fair protocols.
\newblock \emph{In Annual Cryptology Conference}, \penalty0 (1):\penalty0
  421–--439, 2014.

\bibitem[Bradford et~al.(2008)Bradford, Park, Rothkopf, and Park]{brad2008}
Phillip~G. Bradford, Sunju Park, Michael~H Rothkopf, and Heejin Park.
\newblock Protocol completion incentive problems in cryptographic vickrey
  auctions.
\newblock \emph{Electronic Commerce Research}, 8\penalty0 (1):\penalty0 57--77,
  2008.

\bibitem[Buterin et~al.()Buterin, Conner, Dudley, Slipper, and
  Norden]{buterin2019eip1559}
Vitalik Buterin, Eric Conner, Rick Dudley, Matthew Slipper, and Ian Norden.
\newblock Ethereum improvement proposal 1559: Fee market change for eth 1.0
  chain.
\newblock \url{https://github.com/ethereum/EIPs/blob/master/EIPS/eip-1559.md}.

\bibitem[Chen et~al.(2019)Chen, Papadimitriou, and
  Roughgarden]{chen2019axiomatic}
Xi~Chen, Christos Papadimitriou, and Tim Roughgarden.
\newblock An axiomatic approach to block rewards.
\newblock In \emph{Proceedings of the 1st ACM Conference on Advances in
  Financial Technologies}, pages 124--131, 2019.

\bibitem[Chung and Shi(2021)]{chung2021foundations}
Hao Chung and Elaine Shi.
\newblock Foundations of transaction fee mechanism design.
\newblock \emph{arXiv preprint arXiv:2111.03151}, 2021.

\bibitem[Cleve(1986)]{cleve86}
Richard Cleve.
\newblock Limits on the security of coin flips when half the processors are
  faulty.
\newblock \emph{In Proceedings of the eighteenth annual ACM symposium on Theory
  of computing}, \penalty0 (1):\penalty0 364–--369, 1986.

\bibitem[Daskalakis et~al.(2020)Daskalakis, Fishelson, Lucier, Syrgkanis, and
  Velusamy]{daskalakis2020simple}
Constantinos Daskalakis, Maxwell Fishelson, Brendan Lucier, Vasilis Syrgkanis,
  and Santhoshini Velusamy.
\newblock Simple, credible, and approximately-optimal auctions.
\newblock In \emph{Proceedings of the 21st ACM Conference on Economics and
  Computation}, pages 713--713, 2020.

\bibitem[Dolev et~al.(2003)Dolev, Dwork, and Naor]{dolev2003nonmalleable}
Danny Dolev, Cynthia Dwork, and Moni Naor.
\newblock Nonmalleable cryptography.
\newblock \emph{SIAM review}, 45\penalty0 (4):\penalty0 727--784, 2003.

\bibitem[Ferreira et~al.(2021)Ferreira, Moroz, Parkes, and
  Stern]{ferreira2021dynamic}
Matheus V.~X. Ferreira, Daniel~J. Moroz, David~C. Parkes, and Mitchell Stern.
\newblock Dynamic posted-price mechanisms for the blockchain transaction-fee
  market.
\newblock In \emph{Proceedings of the 3rd ACM conference on Advances in
  Financial Technologies}, AFT '21, New York, NY, USA, 2021. Association for
  Computing Machinery.

\bibitem[Ferreira and Weinberg(2020)]{ferreira2020credible}
Matheus~VX Ferreira and S~Matthew Weinberg.
\newblock Credible, truthful, and two-round (optimal) auctions via
  cryptographic commitments.
\newblock In \emph{Proceedings of the 21st ACM Conference on Economics and
  Computation}, pages 683--712, 2020.

\bibitem[Ferreira and Weinberg(2021)]{ferreira2021proof}
Matheus~VX Ferreira and S~Matthew Weinberg.
\newblock Proof-of-stake mining games with perfect randomness.
\newblock In \emph{Proceedings of the 22nd ACM Conference on Economics and
  Computation}, pages 433--453, 2021.

\bibitem[Li(2017)]{li2017obviously}
Shengwu Li.
\newblock Obviously strategy-proof mechanisms.
\newblock \emph{American Economic Review}, 107\penalty0 (11):\penalty0
  3257--87, 2017.

\bibitem[Liu et~al.(2019)Liu, Mierendorff, Shi, and Zhong]{zhong2019}
Qingmin Liu, Konrad Mierendorff, Xianwen Shi, and Weijie Zhong.
\newblock Auctions with limited commitment.
\newblock \emph{American Economic Review}, \penalty0 (109):\penalty0 876–910,
  2019.

\bibitem[Myerson(1981)]{myerson1981optimal}
Roger~B Myerson.
\newblock Optimal auction design.
\newblock \emph{Mathematics of operations research}, 6\penalty0 (1):\penalty0
  58--73, 1981.

\bibitem[Nakamoto(2009)]{nakamoto2008bitcoin}
Satoshi Nakamoto.
\newblock Bitcoin: A peer-to-peer electronic cash system, 2009.
\newblock URL \url{http://www.bitcoin.org/bitcoin.pdf}.

\bibitem[Pycia and Raghavan(2021)]{raghavan2021}
Marek Pycia and Madhav Raghavan.
\newblock Non-bossiness and first-price auctions.
\newblock 2021.

\bibitem[Reijsbergen et~al.(2021)Reijsbergen, Sridhar, Monnot, Leonardos,
  Skoulakis, and Piliouras]{reijsbergen2021transaction}
Dani{\"e}l Reijsbergen, Shyam Sridhar, Barnab{\'e} Monnot, Stefanos Leonardos,
  Stratis Skoulakis, and Georgios Piliouras.
\newblock Transaction fees on a honeymoon: Ethereum's eip-1559 one month later.
\newblock \emph{arXiv preprint arXiv:2110.04753}, 2021.

\bibitem[Roughgarden(2021)]{roughgarden2021transaction}
Tim Roughgarden.
\newblock Transaction fee mechanism design.
\newblock \emph{arXiv preprint arXiv:2106.01340}, 2021.

\bibitem[Skreta(2013)]{skreta2013}
Vasiliki Skreta.
\newblock Optimal auction design under non-commitment.
\newblock \emph{Journal of Economic Theory}, \penalty0 (159):\penalty0 854–
  890, 2013.

\end{thebibliography}

\appendix

\section{Omitted Proofs From Section~\ref{sec:adra}}\label{app:adra-proofs}

\begin{claim}\label{claim:revenue-inequality}
Let $(G, \bm S)$ be the protocol resulting from the second-price auction with reserve $r$ where bidder $i \in [n]$ bids $b_i = v_i$. Let $u(\bm v)$ be the second highest bid among $\bm v = (v_1, v_2, \ldots, v_n).$ Then
$$\rev^{(G, \bm S)}(\bm D) = \e{u(\bm v) \cdot \ind{u(\bm v) \geq r}} + \e{r \cdot \ind{\max_{i \in [n]} v_i \geq r, u(\bm v) < r}}.$$
\end{claim}
\begin{proof}
In the second-price auction with reserve $r$, the highest bidder pays the maximum between $r$ and the second highest bidder. Thus the revenue is
\begin{align*}
    \e{\max\{u(\bm v), r\} \cdot \ind{\max_{i \in [n]} v_i \geq r}} = \e{r \cdot \ind{\max_{i \in [n]} v_i \geq r, u(\bm v) < r}} + \e{u(\bm v) \cdot \ind{u(\bm v) \geq r}}
\end{align*}
\end{proof}
\begin{proof}[Proof of Lemma~\ref{lem:apa-lowerbound}]
Let $u(\bm v)$ denote the second-highest bid in bid profile $\bm v$. Recall the ascending price auction terminates once the second-highest bidder quits. Since bidders start bidding the reserve price $r$, the expected number of rounds for the ascending price auction is
\begin{align*}
&\Theta(1) + \e{\frac{u(\bm v)-r}{\varepsilon} \cdot \ind{u(\bm v) \geq r}}\\
&\qquad = \Theta(1) + \frac{\rev^{(G, \bm S)}(\bm D) - r\pr{\max_{i \in [n]} v_i \geq r, u(\bm v) < r} - r\pr{u(\bm v) \geq r}}{\varepsilon}\\
&\qquad = \Theta(1) + \frac{\rev^{(G, S)}(\bm D) - r\pr{\max_{i \in [n] v_i \geq r}}}{\varepsilon}
\end{align*}
where the first equality is from Claim~\ref{claim:revenue-inequality}. The second equality observes $u(\bm v) \geq r$ implies $\max_{i \in [n]} v_i \geq r$. 
\end{proof}

\begin{proof}[Omitted proofs from Example~\ref{example:equal-revenue}]
Consider $n$ bidders with valuations drawn i.i.d. from the equal revenue distribution where $\pr{v_i \geq x} = \frac{1}{x}$ for all $x \geq 1$. Note the virtual value of this distribution is $\vvalue(x) = 0$ for all $x$. Thus the optimal auction sets a reserve price of $1$, sells the item to the highest bidder and charges the second highest bid. To compute the revenue, note that for each fixed $i$, if we conditioned on $v_i$ being the highest bid, then the payment of bidder $i$ is the maximum value among bidders $j \neq i$. Therefore,
\begin{align*}
    \rev(\bm D) & = n \cdot \e{\max_{j \neq i}\{ v_j \} \cdot \ind{v_i >  \max_{j \neq i} \{ v_j \}}}\\
    &= n \e{\max_{j \neq i}\{ v_j \} \cdot \pr{v_i > \max_{j \neq i} \{v_j\} | \bm v_{-i}} }\\
    &= n \e{\max_{j \neq i}\{ v_j \} \cdot \frac{1}{\max_{j \neq i}\{ v_j \}}}\\
    &= n
\end{align*}
as desired.
\end{proof}

%
\begin{proof}[Proof of Proposition~\ref{prop:adra-nash}]
We will require the following definition about the view of bidder $i$ during the execution of a safe deviation. We use a subscript $i$ to denote the state of the auction in the view of bidder $i$. For example, $n_i$ denotes the number of bidders bidder $i$ (believes) are in the auction. Then $n_i$ might be different from $n_j$ for bidder $j \neq i$.
\begin{definition}[View of bidder $i$]
Let $(b_i^1, \ldots, b_i^{i-1}, b_i^i, b_i^{i+1}, \ldots, b_i^{n_i})$ denote the bids committed to bidder $i$---bidder $i$ believes there is $n_i$ bidders in the auction. Let $\ell_i^j$, for $j \in [n_i]$, be the level where, according to the auctioneer, bidder $j$ quits, aborts or becomes the only remaining bidder in the auction. Let $b_i = b_i^i$ the bid of (real) bidder $i$ and $\ell_i = \ell_i^i$ the level bidder $i$ quits or become the only bidder left. Note that $\bm b_i$---the bids sent to bidder $i$---is a function of all values except $v_i$ since bidder $i$ has no yet revealed $v_i$, but it is possible the auctioneer learned $\bm v_{-i}$ before first interacting with bidder $i$. However, the level bid $b_i^j$ quits---$\bm \ell_i$---is a function of all values. We say {\em bidder $j \in [n_i]$ aborts to bidder $i \in [n]$} if, according to the auctioneer, bidder $j$ aborted the auction at level $\ell_i^j$. We say the {\em auctioneer aborts bid $b_i^j$} if, according to the auctioneer, bidder $j \in [n_i]$ aborted from the auction, but bidder $j$ is not a real bidder that aborted from the auction
\end{definition}
Note that if the auctioneer implements a safe deviation and allocates the item to bidder $i$, but aborts $b_i^j$ at level $\ell_i^j$, then bidder $i$ expects to receive payment $g^{-1}(\ell_i^j)$ from bidder $j$'s deposit which must come from the auctioneer himself. Moreover, if bidder $i$ receives the item, the auctioneer receives a payment of at most $g^{-1}(\ell_i)$. This implies the auctioneer does not benefit from aborting a fake bid $b_i^j$ when $\ell_i^j \geq \ell_i$:
\begin{proposition}\label{prop:auctioneer-strategy}
There is an optimal safe deviation of ADRA where $\ell_i^j < \ell_i$ whenever the auctioneer aborts $b_i^j$.
\end{proposition}
\begin{proof}
Consider a safe deviation $(G', \bm S^{ADRA})$ where the auctioneer aborts $b_i^j$ at level $\ell_i^j \geq \ell_i$. The auctioneer only receives a payment from bidder $i$ only if they win the item and pay at most $g^{-1}(\ell_i)$. However, the auctioneer pays $g^{-1}(\ell_i^j) \geq g^{-1}(\ell_i)$ to bidder $i$ for aborting $b_i^j$. Thus the utility of the auctioneer is non-positive. At the time the auctioneer aborts $b_i^j$, the auctioneer had the option to {\em not} abort $b_i^j$ in the view of bidder $i$ which weakly dominates aborting $b_i^j$. This proves there is a safe deviation $(G'', \bm S^{ADRA})$ that weakly dominates $(G', \bm S^{ADRA})$ where the auctioneer does not abort $b_i^j$ as desired.
\end{proof}
Let $G'$ be a safe deviation from ADRA. To show $(G', \bm S^{ADRA}(\bm v))$ is an {\em ex post} Nash equilibrium, we will show that for all $i \in [n]$, for all $\bm v$, picking $b_i = v_i$ weakly dominates any other strategy. Observation~\ref{obs:bidder-strategy} pins down most of the strategy space for bidder $i$: we only need to consider the case where bidder $i$ plays $S_i^{ADRA}(b_i)$ for some $b_i$ (not necessarily equal to $v_i$).

Let $\ell_i(\bm b)$ denote the highest level played by bidder $i$ when bidder $j \in [n]$ bids $b_j$. For the proof, we will analyse the following cases:

\begin{enumerate}
\item Consider the case where $\ell_i(b_i, \bm v_{-i}) < \ell_i(\bm v)$. Note $\ell_i(b_i, \bm v_{-i})$ is the last level bidder $i$ plays either because bidder $i$ quits or bidder $i$ becomes the last bidder in the auction. However, the fact $\ell_i(\bm v) > \ell_i(b_i, \bm v_{-i})$ implies bidder $i$ quits at level $\ell_i(b_i, \bm v_{-i})$ since the auctioneer could distinguish---provided our crytographic scheme is perfectly hiding---the case where bidder $i$ bids $v_i$ from the case where bidder $i$ bids $b_i$. To be concrete: the information set $I_i$ where bidder $i$ quits at level $\ell_i(b_i, \bm v_{-i})$ is the first where the auctioneer can distinguish the case where bidder $i$ bids $v_i$ from $b_i$ (and all actions of the auctioneer and bidder $j \neq i$ up to this point are identical). If the auctioneer did not end the auction at level $\ell_i(b_i, \bm v_{-i})$ when bidder $i$ bid $v_i$ (and choose not to quit), then in the view of bidder $i$, there is at least another bid $b_i^j$ that also did not quit at level $\ell_i(b_i, \bm v_{-i})$. If $j$ is real bidder, bidder $j$ never quits. If $j$ was a fake bidder, from Proposition~\ref{prop:auctioneer-strategy}, the auctioneer does not quit $b_i^j > b_i$ once bidder $i$ quits at level $\ell_i(b_i, \bm v_{-i})$. Therefore, bidder $i$ does not receive the item when they bid $b_i$ since the auctioneer later reveals a higher bid $b_i^j$. This proves bidding $v_i$ weakly dominates bidding $b_i$.

\item Consider the case where $\ell_i(b_i, \bm v_{-i}) = \ell_i(\bm v)$. The security of our cryptographic commitment ensures the auctioneer cannot distinguish the case where bidder $i$ bids $b_i$ from the case where they bid $v_i$ until the bid is revealed. From Proposition~\ref{prop:auctioneer-strategy}, the auctioneer aborts no bids to bidder $i$ once bidder $i$ quits. Therefore, the set of bids revealed to bidder $i$ when they bid $b_i$ is identical to the set of bids revealed when they bid $v_i$. Since the auctioneer allocates the item to the highest bidder which pays the second-highest bid, bidder $i$ can only improve their utility if they did not receive the item when they bid $v_i$, but receive the item when they bid $b_i$. That implies $b_i > v_i$ and bidder $i$ pays more than they value the item. This proves bidding $v_i$ weakly dominates bidding $b_i$.

\item Consider the case $\ell_i(b_i, \bm v_{-i}) > \ell_i(\bm v)$. Assuming our cryptographic scheme is perfectly hiding, the auctioneer cannot distinguish the case where bidder $i$ bids $b_i$ from the case where bidder $i$ bids $v_i$, the fact $\ell_i(\bm v) < \ell_i(b_i, \bm v_{-i})$ implies bidder $i$ tentatively quits at level $\ell_i(\bm v)$ when they bid $v_i$---$g(v_i) = \ell_i(\bm v)$. By definition of $S_i^{ADRA}$, bidder $i$ would quit at level $g(b_i)$; therefore, $\ell_i(b_i, \bm v_{-i}) \leq g(b_i)$. Combined with the fact $\ell_i(b_i, \bm v_{-i}) > \ell_i(\bm v) = g(v_i)$, we conclude $g(b_i) > g(v_i)$. Thus $b_i > v_i$ and if bidder $i$ wins the item, they pay at least $g^{-1}(b_i) \geq v_i$ resulting in nonpositive utility. Thus bidding $v_i$ weakly dominates bidding $b_i$.
\end{enumerate}
This covers all the cases and proves the bidding $v_i$ weakly dominates bidding $b_i \neq v_i$. Our proof only assumes the auctioneer implements a safe deviation (or the honest auction) which implies $(G', \bm S^{ADRA})$ forms an {\em ex post} Nash equilibrium as desired.
\end{proof}

\section{Omitted Proofs From Section~\ref{sec:communication}}\label{app:regular-proofs}


\begin{proof}[Proof of Theorem~\ref{thm:adra-upperbound}]
Recall that there are $n$ bidders drawn i.i.d. from $D$, the reserve price is $r$, and the level function is
$$g(b) = \lceil\log_{1+\varepsilon}\left(\frac{b}{r}\right)\rceil.$$
By assumption, bidder $i$ follows the strategy $S_i^{ADRA}(v_i)$ which implies bidder $i$ quits at the first level $k$ where $k \geq g(v_i)$. Let $u(\bm v)$ be the second-highest bid in the bid profile $\bm v$. Then the communication complexity is proportional to the number of rounds it takes until the second-highest bidder quits:
\begin{align*}
    O(1) + \e{\log_{1+\varepsilon}\left(\frac{u(\bm v)}{r}\right) \cdot \ind{u(\bm v) \geq r}}
\end{align*}
From Jensen's inequality,
\begin{align*}
    &\e{\log_{1+\varepsilon}\left(\frac{u(\bm v)}{r}\right) \cdot \ind{u(\bm v) \geq r}} \leq \pr{u(\bm v) \geq r} \log_{1+\varepsilon}\e{\frac{u(\bm v)}{r} | u(\bm v) \geq r}\\
    &\qquad= \pr{u(\bm v) \geq r} \log_{1+\varepsilon} \frac{\e{u(\bm v) \cdot \ind{u(\bm v) \geq r}}}{r \cdot \pr{u(\bm v) \geq r}}\\
    &\qquad= \pr{u(\bm v) \geq r} \log_{1+\varepsilon} \frac{\e{u(\bm v) \cdot \ind{u(\bm v) \geq r}}}{r} + \pr{u(\bm v) \geq r}\log \frac{1}{\pr{u(\bm v) \geq r}}\\
    &\qquad\leq \log_{1+\varepsilon} \frac{\e{u(\bm v) \cdot \ind{u(\bm v) \geq r}}}{r} + \frac{1}{e} \quad \text{$\max_{x \in \mathbb R} x\log(1/x) \leq 1/e$,}\\
    &\qquad\leq \log_{1+\varepsilon} \frac{\rev^{(G, \bm S)}(D)}{r} + \frac{1}{e} \quad \text{From Claim~\ref{claim:revenue-inequality}.}
\end{align*}
The chain of inequalities witnesses that the communication complexity is at most $O(1) + \log_{1+\varepsilon} \frac{\rev^{(G, \bm S)}(D)}{r}$ as desired.
\end{proof}


\begin{proof}[Proof of Theorem~\ref{thm:regular}]
Recall $F$ is the cumulative density function of a regular distribution $D$ and $x \coloneqq \inf\{p : \pr[v \leftarrow D]{v \geq p} \leq \frac{1}{n}\}$. We first define and prove the following four claims:
\begin{claim}\label{claim:high-reserve}
If Myerson reserve $r(D) \geq x$, the communication complexity is at most $\frac{1+\varepsilon}{2\varepsilon}$.
\end{claim}
\begin{proof}[Proof of Claim~\ref{claim:high-reserve}]
Let $u(\bm v)$ be the second highest bid in $\bm v$ and let $k$ denote the level where the second highest bidder quits. For any $z \geq 0$, $k > z$ if and only if the second highest bidder does not quit at level $z$ which implies $g(u(\bm v)) > z$. Thus
\begin{align*}
\pr{k > z} &= \pr{u(\bm v) >  (1+\varepsilon)^z \cdot r}
\end{align*}
The event $u(\bm v) >  (1+\varepsilon)^z \cdot r(D)$ implies there are at least two bidders with value bigger than $(1+\varepsilon)^z \cdot r(D)$. From union bound,
\begin{align*}
\pr{u(\bm v) >  (1+\varepsilon)^z \cdot r(D)} &\leq {n\choose 2} \pr[v \leftarrow D]{v >  (1+\varepsilon)^z \cdot r(D)}^2\\
&= {n\choose 2} \left(\frac{(1+\varepsilon)^z \cdot r(D)}{(1+\varepsilon)^z \cdot r(D)} \cdot \pr[v \leftarrow D]{v > (1+\varepsilon)^z \cdot r(D)} \right)^2
\end{align*}
For Myerson reserve $r(D)$, we have $r(D) \pr[v \leftarrow D]{v \geq r(D)} \geq p \pr[v \leftarrow D]{v \geq p}$ for all $p \in \mathbb R$. Thus
\begin{align*}
\pr{k > z} &\leq {n\choose 2} \left(\frac{r(D)}{(1+\varepsilon)^z \cdot r(D)} \cdot \pr[v \leftarrow D]{v \geq r(D)} \right)^2\\
&= {n\choose 2}  \frac{\pr[v \leftarrow D]{v \geq r}^2 }{(1+\varepsilon)^{2z}}
\end{align*}
We now compute the expected number of rounds and find that:
\begin{align*}
\e{k} = \sum_{z = 0}^\infty \pr{k > z} \leq \sum_{z = 0}^\infty {n \choose 2} \frac{\pr[v \leftarrow D]{v \geq r(D)}^2}{(1+\varepsilon)^{2z}}.
\end{align*}
Recall that $\pr{v \geq r(D)} \leq \pr{v \geq x}$ since $r(D) \geq x$, and $\pr{v \geq x} \leq \frac{1}{n}$ by definition of $x$. Together with the fact that ${n \choose 2} = \frac{n(n-1)}{2} \leq \frac{n^2}{2}$, we have
$$\e{k} \leq \frac{n^2}{2} \pr{v \geq x} \sum_{z = 0}^\infty \frac{1}{(1+\varepsilon)^{2z}} \leq \frac{(1+\varepsilon)^2}{2\varepsilon(2+\varepsilon)} \leq \frac{1+\varepsilon}{2\varepsilon}$$
as desired.
\end{proof}

The case where $r(D) < x$ will require the following property for the revenue curve $p\pr[v \leftarrow D]{v \geq p}$ of regular distributions. It states that the revenue curve is decreasing for $p \geq r(D)$ where $r(D)$ is the Myerson reserve.
\begin{claim}\label{claim:decreasing-revenue}
Let $D$ be a regular distribution. For any $p \geq r(D)$, the revenue curve $p \pr[v \leftarrow D]{v \geq p}$ is nonincreasing in $p$.
\end{claim}
\begin{proof}
Recall that the virtual value function of $D$ is $\phi(p) = p - \frac{1-F(p)}{f(p)}$, which is monotone nondecreasing from the definition of regularity. Then
\begin{align*}
    p \geq r(D) &\iff \phi(p) \geq \phi(r(D)) &\quad \text{From regularity of $D$,}\\
    &\iff \phi(p) \geq 0 &\quad \text{Since $\phi(r(D)) = 0$,}\\
    &\iff f(p) \cdot \phi(x) \geq 0\\
    &\iff -f(p) \cdot \left(p-\frac{1-F(p)}{f(p)}\right) \leq 0 & \quad \text{From definition of $\phi(p)$,}\\
    &\iff (1-F(p)) + p(-f(p)) \leq 0
\end{align*}
Note that $(1-F(p)) - pf(p)$ is the derivative of the revenue curve. The fact that the derivative is nonpositive implies the revenue curve is nonincreasing for $p \geq r(D)$ as desired.
\end{proof}
\begin{claim}\label{claim:ineq-1}
Let $D$ be regular and $p \geq r(D)$. Then for any $\delta \geq 1$, $\pr[v \leftarrow D]{v > \delta p} \leq \frac{\pr[v \leftarrow D]{v > p}}{\delta}$.
\end{claim}
\begin{proof}
From Claim~\ref{claim:decreasing-revenue}, $p\pr[v \leftarrow D]{v > p} \geq p \delta \pr[v \leftarrow D]{v > \delta p}$. Rearranging the inequality proves the claim.
\end{proof}


\begin{claim}\label{claim:low-reserve}
If the Myerson reserve $r(D) < x$, then the communication complexity is $O(\log_{1+\varepsilon}(x/r(D)))$.
\end{claim}
\begin{proof}
Let $t = 2\log_{1+\varepsilon}(x/r(D))$ and note $x = r(D)(1+\varepsilon)^{t/2}$. Define $k$ to be level the auction terminates. Let $u(\bm v)$ denote the second highest bid in $\bm v$. Computing the expected value of $k$ gives:
\begin{align*}
\e{k} &= \sum_{z=0}^{\infty} \pr{k > z}\\
&= \sum_{z=0}^{\infty} \pr{u(\bm v) >  (1+\varepsilon)^z \cdot r}\\
&= \sum_{z=0}^{t} \pr{u(\bm v) > (1+\varepsilon)^z \cdot r} + \sum_{z=t+1}^{\infty} \pr{u(\bm v) > (1+\varepsilon)^z \cdot r}
\end{align*}
The second line observes the auction terminates at a level above $z$ if and only if the second highest bidder quits at a level $z$. Next, we upper bound the two terms separately.

For the first term, $\sum_{z=0}^{t} \pr{u(\bm v) >  (1+\varepsilon)^z \cdot r} \leq t+1$. For the second term, observe that the fact the second highest bidder bids more than $(1+\varepsilon)^z r(D)$ implies there is at least two bidders that bid more than $(1+\varepsilon)^z r(D)$. From union bound,
\begin{align*}
    &\sum_{z = t + 1}^\infty \pr{u(\bm v) > (1+\varepsilon)^z \cdot r(D)}\\
    &\quad \leq {n \choose 2} \sum_{z = t + 1}^\infty \pr[v \leftarrow D]{v > (1+\varepsilon)^z \cdot r(D)}^2 & \quad \text{From union bound,}\\
    &\quad \leq \frac{n^2}{2} \sum_{z = t + 1}^\infty \pr[v \leftarrow D]{v > (1+\varepsilon)^{z/2}(1+\varepsilon)^{z/2} r(D)}^2\\
    &\quad \leq \frac{n^2}{2} \sum_{z = t+1}^\infty \pr[v \leftarrow D]{v > (1+\varepsilon)^{z/2}(1+\varepsilon)^{t/2} r(D)}^2 & \quad \text{Since $t\leq z$,}\\
    &\quad \leq \frac{n^2}{2} \sum_{z = t+1}^\infty  \frac{\pr[v \leftarrow D]{v > (1+\varepsilon)^{t/2} r(D)}^2}{(1+\varepsilon)^{z}} & \quad \text{From Claim~\ref{claim:ineq-1}.}
\end{align*}
Next observe the geometric series $\sum_{z = t+1}^\infty \frac{1}{(1+\varepsilon)^z} = \frac{1}{\varepsilon(1+\varepsilon)^t} \leq \frac{1}{\varepsilon}$. By definition, $t/2 \geq g(x)$ and (from definition of $x$) the probability that a bid exceeds $x$ is at most $\frac{1}{n}$. Therefore, $\pr[v \leftarrow D]{v > (1+\varepsilon)^{t/2} r(D)} \leq \frac{1}{n}$. This proves the second term is at most $\frac{1}{2\varepsilon}$. Combining both terms, we conclude $\e{k} \leq t + 1 + \frac{1}{2\varepsilon}$. This proves the communication complexity is $O\left(\log_{1+\varepsilon}\left(\frac{x}{r(D)}\right)\right)$ as desired.
\end{proof}

We are ready to conclude the proof of Theorem~\ref{thm:regular}. Recall that $x$ is the smallest price such that a value drawn from $D$ exceeds $x$ with probability at most $1/n$. Claim~\ref{claim:high-reserve} states the communication complexity is at most $O\left(\frac{1+\varepsilon}{\varepsilon}\right)$ when $r(D) \geq x$. Claim~\ref{claim:low-reserve} states the communication complexity is at most $O\left(\log_{1+\varepsilon}\left(\frac{x}{r(D)}\right)\right)$ when $r(D) < x$. Therefore, the communication complexity for ADRA is at most the maximum among this two quantities as desired.
\end{proof}


\end{document}